\documentclass[nocopyrightspace]{sigplanconf}

\usepackage[nospace]{cite}

\usepackage{graphicx}
\usepackage{pgf}
\usepackage{amsmath}
\usepackage{amssymb}
\usepackage{array}
\usepackage{url}

\usepackage{listings}
\usepackage[english]{babel}
\usepackage[utf8]{inputenc}
\usepackage{verbatim}
\graphicspath{{figures/}}

\makeatletter
\newcommand{\FIXME}[2]{\@latex@warning{FIXME: #1}%
  {\raggedright\quad\sffamily\slshape[~#1: #2~]\quad}}
\makeatother

\definecolor{lightgray}{rgb}{.94,.94,.94}
\definecolor{darkgreen}{rgb}{0,.6,0}
\definecolor{darkblue}{rgb}{0,0,.6}

\newcommand{\OCCL}{{\color{darkgreen}[\![}}
\newcommand{\OCCR}{{\color{darkgreen}]\!]}}

\lstdefinelanguage{custom}{%
  keywords={[1]for,do,while,if,else,break,continue,return},
  keywords={[2]const,volatile,static,signed,unsigned,%
    void,char,short,long,int,float,double,boolean,%
    recording,view,size_t,rec_t},
  literate={[[}{{\OCCL}}1 {]]}{{\OCCR}}1,
  string=[b]",
  comment=[l]//,
  morecomment=[s]{/*}{*/},
  mathescape=true,
  flexiblecolumns=true,
  tabsize=2,
  captionpos=b,
  frame=single,
  framerule=0pt,
  aboveskip=1pt,
  belowskip=1pt,
  framesep=1pt,
  basicstyle=\ttfamily\scriptsize,
  keywordstyle={[1]\color{darkblue}},
  keywordstyle={[2]\color{darkgreen}},
  keywordstyle={[3]\color{orange}},
  keywordstyle={[4]\color{red}},
  emphstyle=\slshape,
  identifierstyle=\color{black},
  commentstyle=\color{darkgray},
  stringstyle=\color{green}
}

\title{The Potential of Synergistic Static, Dynamic and Speculative
  Loop Nest Optimizations for Automatic Parallelization}

\authorinfo{Riyadh Baghdadi$^1$, Albert Cohen$^1$, \\
  Cedric Bastoul$^1$, Louis-Noël Pouchet$^2$ and Lawrence Rauchwerger$^3$ \\
  \small
  $^1$ INRIA Saclay and LRI, Paris-Sud 11 University \\
  \small
  $^2$ The Ohio State University
  \small
  $^3$ Dept.\ of Computer Science and Engineering, Texas A\&M University}{}{}

\begin{document}

\maketitle


\section{Introduction}
\label{sec:introduction}

Research in automatic parallelization of loop-centric programs started
with static analysis, then broadened its arsenal to include dynamic
inspection-execution and speculative execution, the best results
involving hybrid static-dynamic schemes. Beyond the detection of
parallelism in a sequential program, scalable parallelization on
many-core processors involves hard and interesting parallelism
adaptation and mapping challenges. These challenges include tailoring
data locality to the memory hierarchy, structuring independent tasks
hierarchically to exploit multiple levels of parallelism, tuning the
synchronization grain, balancing the execution load, decoupling the
execution into thread-level pipelines, and leveraging heterogeneous
hardware with specialized accelerators.

The polyhedral framework allows to model, construct and apply very
complex loop nest transformations addressing most of the parallelism
adaptation and mapping challenges.  But apart from hardware-specific,
back-end oriented transformations (if-conversion, trace scheduling,
value prediction), loop nest optimization has essentially ignored
dynamic and speculative techniques.  Research in polyhedral
compilation recently reached a significant milestone towards the
support of dynamic, data-dependent control flow. This opens a large
avenue for blending dynamic analyses and speculative techniques with
advanced loop nest optimizations.  Selecting real-world examples from
SPEC benchmarks and numerical kernels, we make a case for the design
of synergistic static, dynamic and speculative loop transformation
techniques.  We also sketch the embedding of dynamic information,
including speculative assumptions, in the heart of affine
transformation search spaces.


\section{Experimental Study}
\label{sec:experiments}

We consider four motivating benchmarks, illustrating three
combinations of dynamic analyses and loop transformations.

Our experiments target three multicore platforms:
\begin{itemize}
\item 2-socket quad-core Intel Xeon E5430, 2.66GHz, 16GB RAM --- 8 cores;
\item 4-socket quad-core AMD Opteron 8380, 2.50GHz, 64GB RAM --- 16 cores;
\item 4-socket hexa-core Intel Xeon E7450, 2.40GHz, 64GB RAM --- 24 cores.
\end{itemize}
We use OpenMP as the target of automatic and manual
transformations. Baseline and optimized codes were compiled with
Intel's compiler ICC 11.0, with options \texttt{-fast -parallel -openmp}.

\subsection{Dynamic techniques may be neither
  necessary nor profitable}

The SPEC CPU2000 \textsf{183.equake} and \textsf{179.art} benchmarks
have frequently been used to motivate dynamic parallelization
techniques. We show that static transformation and parallelization
techniques can easily be extended to handle the limited degree of
data-dependent behavior in these programs.

Figure~\ref{fig:equake} shows the \texttt{smvp()} function of
\textsf{equake}, well known for its ``sparse'' reduction pattern (a
histogram computation).  The value of \texttt{col} is read from an
array; it is not possible to establish at compilation time whether and
when dependences will occur upon accumulating on
\texttt{w[col][0]}. Zhuang et al.\ \cite{zhuang_exploiting_2009} used
automatically generated inspection slices to parallelize this
loop. The inspector slice is a simplified version of the original loop
to anticipate the detection of dynamic dependences. In the case of
\textsf{equake}, it computes the values of \texttt{col} within a
sliding window of loop iterations to detect possible conflicts and
build a safe schedule at run-time.

Speculation has also been used to handle unpredictable memory accesses
in \textsf{equake}. Oancea et al.\ \cite{oancea_lightweight_2009}
implemented a speculative system to spot conflicts at runtime. When a
thread detects a dependence violation, it kills other speculative
threads and rolls back. If the number of rollbacks exceeds 1\%, the
execution proceeds in serial mode. This approach is similar to
\cite{mehrara_parallelizing_2009} which uses transactional memory to
implement thread-level speculation to parallelize \textsf{equake}.
Speculation is an interesting solution for dynamic parallelization,
but has a high overhead due to memory access tracing, dependence
checking, rollback and/or commit overhead.

Interestingly, in the case of \textsf{equake}, one may avoid
inspection and speculation altogether. It is sufficient to enforce
atomic execution of the sparse reduction to \texttt{w[col][0]}. This
can be done with hardware atomic instructions. An alternative is to
privatize the \texttt{w} array to implement a conflict-free parallel
reduction. This induces some overhead to scan the private arrays (as
many as concurrent threads) and sum up the partial accumulation
results.

In the case of \textsf{art}, atomic execution of the tailing part of
the \texttt{match()} function is also sufficient to make an outer loop
parallel, see Figure~\ref{fig:art}. Since we are also dealing with a
reduction, the privatization alternative applies as well.

\begin{figure}[h!tb]
  \begin{lstlisting}
for (i=0; i<nodes; i++) {
  Anext = Aindex[i];
  Alast = Aindex[i+1];

  sum0 = A[Anext][0][0]*v[i][0]+
    A[Anext][0][1]*v[i][1]+
    A[Anext][0][2]*v[i][2];

  Anext++;
  while (Anext<Alast) {
    col = Acol[Anext];

    sum0 += A[Anext][0][0]*v[col][0]+
      A[Anext][0][1]*v[col][1]+
      A[Anext][0][2]*v[col][2];

    // Sparse reduction
    w[col][0] += A[Anext][0][0]*v[i][0]+
      A[Anext][1][0]*v[i][1]+
      A[Anext][2][0]*v[i][2];

    Anext++;
  }
  w[i][0] += sum0;
}
  \end{lstlisting}

  \caption{\textsf{equake}, core of the \texttt{smvp()} function}
  \label{fig:equake}
\end{figure}

\begin{figure}[h!tb]
  \begin{lstlisting}
if (match_confidence > highest_confidence[winner]) {
  highest_confidence[winner] = match_confidence;
  set_high[winner] = TRUE;
}
  \end{lstlisting}

  \caption{\textsf{art}, end of the \texttt{match()} function}
  \label{fig:art}
\end{figure}

Figure~\ref{fig:speedup_equake_art} compares the speedup results of
static loop transformation vs.\ speculative conflict management with
Intel's McRT Software Transactional Memory (STM)
\cite{saha_mcrt-stm:high_2006}. We run the full benchmark programs on
their \textsf{ref} dataset.  For \textsf{equake}, the static version
uses a hardware atomic instruction version. The STM version fails to
deliver any speedup while the version with hardware atomic
instructions scales reasonably well.\footnote{As already pointed out
  in \cite{cascaval_software_2008}.}  For \textsf{art}, the static
version uses privatization. The critical section is executed rarely
and the grain of parallelism is much bigger, which allows the STM
version to yield some speedups although the statically privatized
version still performs better.

We also conducted experiments with different datasets. In the case of
\textsf{equake}, it has a tremendous impact on the relative
performance of the static privatization and hardware atomic versions,
as shown in Figure~\ref{fig:equake_variants}. With the smaller
\textsf{train} dataset, the privatization version outperforms the
hardware atomic version because the private arrays fit in the cache
and privatization removes all contention on the hardware atomic
instructions. As a side effect, this result advocates for an adaptive
compilation scheme generating multiple versions and a decision tree to
dynamically select the most appropriate version depending on program
behavior and/or on features of the input data.

For both benchmarks, we expect more complex loop transformations like
loop tiling to further improve scalability. But we could not yet find
a tool to automate the process. Instead of pursuing the manual
transformation exploration, we prefer to test this hypothesis on
another set of benchmarks more amenable to automatic parallelization
with classical loop transformation tools. This will be the subject of
the next section.

\begin{figure}[h!tb]
  \begin{tabular}{|l||r|r|r||r|r|r|}
    \cline{2-7}
    \multicolumn{1}{l|}{} &
    \multicolumn{3}{c||}{Static only} &
    \multicolumn{3}{c|}{STM} \\
    \multicolumn{1}{l|}{} &
    8 & 16 & 24 & 8 & 16 & 24 \\
    \cline{2-7}
    \noalign{\vspace{2pt}}
    \hline
    \textsf{equake} &
    2.71 &
    6.51 &
    7.18 &
    0.22 &
    0.34 &
    0.24 \\
    \hline
    \textsf{art} &
    3.69 &
    4.29 &
    4.26 &
    3.60 &
    3.69 &
    3.98 \\
    \hline
  \end{tabular}
  
  \caption{Speedups for \textsf{equake} and \textsf{art}}
  \label{fig:speedup_equake_art}
\end{figure}

\begin{figure}[h!tb]
  \begin{tabular}{|l||r|r|r||r|r|r|}
    \cline{2-7}
    \multicolumn{1}{l|}{} &
    \multicolumn{3}{c||}{\textsf{train(small)}} &
    \multicolumn{3}{c|}{\textsf{ref}} \\
    \multicolumn{1}{l|}{} &
    8 & 16 & 24 & 8 & 16 & 24 \\
    \cline{2-7}
    \noalign{\vspace{2pt}}
    \hline
    Locks &
    0.15 &
    0.21 &
    0.09 &
    0.17 &
    0.22 &
    0.12 \\
    \hline
    STM &
    0.21 &
    0.27 &
    0.16 &
    0.22 &
    0.34 &
    0.24 \\
    \hline
    HW atomic &
    3.18 &
    5.84 &
    5.73 &
    2.71 &
    6.51 &
    7.18 \\
    \hline
    Privatization &
    5.13 &
    6.32 &
    6.01 &
    1.48 &
    2.78 &
    2.04 \\
    \hline
  \end{tabular}

  \caption{Speedups for variants of \textsf{equake} on
    \textsf{train} and \textsf{ref} datasets}
  \label{fig:equake_variants}
\end{figure}

\subsection{Complex loop transformations on data-dependent
  control flow}

Dynamic inspection and speculation are very appropriate for
\texttt{equake}, but we showed that it is not strictly needed to
resort to dynamic analysis to achieve good performance. Moreover it is
not always possible to generate lightweight instrumentation slices or
profitable speculation schemes in general
\cite{rauchwerger_run-time_1998}. A good example where we can not
extract an efficient inspector is the \textsf{Givens} rotation kernel
in Figure~\ref{fig:givens}. It features two nested data-dependent
conditions to distinguish between different complex sine/cosine
computations. These conditions prevent optimization and
parallelization in classical frameworks restricted to affine
conditional expressions and loop bounds \cite{Gir06,Bon08b}.

There are loop-carried dependences from and to all three conditional
branches. These are flow dependences and cannot be eliminated by array
expansion (privatization, renaming) techniques. No dynamic parallelism
detection method alone can find scalable parallelism on this example:
it may only extract parallelism in the inner loops (which can also be
extracted with static techniques, but does not bring significant
performance benefit).  The loop nest must be transformed to express
coarser grain parallelism at the outer loop level. This is the
specialty of affine transformations in the polyhedral framework: here, a
composition of privatization, loop skewing and loop tiling would be
possible \cite{Gir06}. The question is how to automate this
transformation.

Fortunately, profiling the kernel shows that the third (\texttt{else})
branch is almost always executed on dense matrices.  With the
assumption that the third branch is almost always executed, one may
speculatively ignore the dependences arising from the first two
branches. Even better, one may virtually eliminate the \texttt{if}
conditionals from the loop nest, yielding a static control loop
nest. With these assumptions \textsc{Pluto} \cite{Bon08b} is able to
tile the loop nest, which greatly enhance the scalability of the
parallelization. The first part of result is shown in
Figure~\ref{fig:givens_optimized}.\footnote{\texttt{floord(n, d)} and
  \texttt{ceild(n, d)} implement $n/d$ and $(n+d-1)/d$ respectively,
  where $/$ is the Euclidian division and not the truncating integer
  division of C and most ISAs.} As announced earlier, it is more
complex than tiling: the loop nest also needs to be skewed to allow
the outer loops to be permuted. This transformation is always correct,
even when the control flow takes one of the two cold branches. It
happens to preserve the dependences arising from the two cold branches
as well. We are in an ideal situation where the speculative assumption
offers extra flexibility in applying complex transformations, but does
not incur any runtime overhead. The end result is a $7.02\times$
speedup on 8 cores for $5000\times5000$ matrices, see
Figure~\ref{fig:speedup_givens_gauss}.

Interestingly, the fact that dependences are compatible with a
composition and loop skewing and loop tiling can also be captured with
conservative, purely static methods, as demonstrated by
Benabderrahmane et al.\ \cite{Ben10}, resulting in the exact same
code.

\begin{figure}[h!tb]
  \begin{lstlisting}
for (k=0; k<N; k++) {
  for (i=0; i<M-1-k; i++) {
    if (A_r[i+1][k] == 0.0 && A_i[i+1][k] == 0.0) {
      // Data-dependent condition, rarely executed
      for (j=k; j<N; j++) {
        t1_r = A_r[i+1][j];
        t1_i = A_i[i+1][j];
        t2_r = A_r[i][j];
        t2_i = A_i[i][j];
        A_r[i][j] = t1_r;
        A_i[i][j] = t1_i;
        A_r[i+1][j] = t2_r;
        A_i[i+1][j] = t2_i;
      }
    } else if (A_r[i][k] == 0.0 && A_i[i][k] == 0.0) {
      // Data-dependent condition, rarely executed
      ng = sqrt(A_r[i+1][k]*A_r[i+1][k]
        + A_i[i+1][k]*A_i[i+1][k]);
      s_r = A_r[i+1][k] / ng;
      s_i = -A_i[i+1][k] / ng;
      for (j=k; j<N; j++) {
        t1_r = -s_r*A_r[i][j] - s_i*A_i[i][j];
        t1_i = -s_r*A_i[i][j] + s_i*A_r[i][j];
        t2_r =  s_r*A_r[i+1][j] - s_i*A_i[i+1][j];
        t2_i =  s_r*A_i[i+1][j] + s_i*A_r[i+1][j];
        A_r[i][j] = t1_r;
        A_i[i][j] = t1_i;
        A_r[i+1][j] = t2_r;
        A_i[i+1][j] = t2_i;
      }
    } else {
      // Most frequently executed case
      nm = sqrt(A_r[i][k] * A_r[i][k] + A_i[i][k] * A_i[i][k] +
        A_r[i+1][k] * A_r[i+1][k] + A_i[i+1][k] * A_i[i+1][k]);
      nf = sqrt(A_r[i][k] * A_r[i][k] + A_i[i][k] * A_i[i][k]);
      sig_r = A_r[i][k] / nf;
      sig_i = A_i[i][k] / nf;
      c_r = nf / nm;
      s_r = (sig_r * A_r[i+1][k] + sig_i * A_i[i+1][k]) / nm;
      s_i = (sig_i * A_r[i+1][k] - sig_r * -A_i[i+1][k]) / nm;
      for (j=k; j<N; j++) {
        t1_r = -s_r*A_r[i][j] - s_i*A_i[i][j] + c_r*A_r[i+1][j];
        t1_i = -s_r*A_i[i][j] + s_i*A_r[i][j] + c_r*A_i[i+1][j];
        t2_r = c_r*A_r[i][j] + s_r*A_r[i+1][j] - s_i*A_i[i+1][j];
        t2_i = c_r*A_i[i][j] + s_r*A_i[i+1][j] + s_i*A_r[i+1][j];
        A_r[i][j] = t1_r;
        A_i[i][j] = t1_i;
        A_r[i+1][j] = t2_r;
        A_i[i+1][j] = t2_i;
      }
    }
  }
}
  \end{lstlisting}

  \caption{\textsf{Givens} kernel}
  \label{fig:givens}
\end{figure}

\begin{figure}[h!tb]
  \begin{lstlisting}
// Skewed and tiled outer loops
for (c0=-1; c0<=min(floord(M-2, 16), floord(N+M-3, 32)); c0++) {
  lb1 = max(max(max(0, ceild(32*c0-M+2,32)),
    ceild(32*c0-N+1, 32)), ceild(32*c0-31,64));
  ub1 = min(floord(M-2, 32), floord(32*c0+31, 32));

  // Parallel loop on coarse-grain blocks
  #pragma omp parallel for shared(c0,lb1,ub1) \
                           private(c1,c2,c3,c4,cond1,cond2)
  for (c1=lb1; c1<=ub1; c1++) {
    if (c0 <= c1) {
      for (c3=max(0,32*c1);c3<=min(M-2,32*c1+31);c3++) {
        cond1 = (A_r[c3+1][0] == 0.0 && A_i[c3+1][0] == 0.0);
        cond2 = (A_r[c3][0] == 0.0 && A_i[c3][0] == 0.0);
        if (cond1) {
          for (c4=0;c4<=N-1;c4++) {
            t1_r = A_r[c3+1][c4];
            t1_i = A_i[c3+1][c4];
            t2_r = A_r[c3][c4];
            t2_i = A_i[c3][c4];
            A_r[c3][c4] = t1_r;
            A_i[c3][c4] = t1_i;
            A_r[c3+1][c4] = t2_r;
            A_i[c3+1][c4] = t2_i;
          }
        } else if (cond2) {
          ng = sqrt(A_r[c3+1][0]*A_r[c3+1][0]
            + A_i[c3+1][0]*A_i[c3+1][0]);
          s_r = A_r[c3+1][0] / ng;
          s_i = -A_i[c3+1][0] / ng;
          for (c4=0;c4<=N-1;c4++) {
            t1_r = -s_r*A_r[c3][c4] - s_i*A_i[c3][c4];
            t1_i = -s_r*A_i[c3][c4] + s_i*A_r[c3][c4];
            t2_r = s_r*A_r[c3+1][c4] - s_i*A_i[c3+1][c4];
            t2_i = s_r*A_i[c3+1][c4] + s_i*A_r[c3+1][c4];
            A_r[c3][c4] = t1_r;
            A_i[c3][c4] = t1_i;
            A_r[c3+1][c4] = t2_r;
            A_i[c3+1][c4] = t2_i;
          }
        } else {
          nm = sqrt(A_r[c3][0]*A_r[c3][0]
            + A_i[c3][0]*A_i[c3][0]
            + A_r[c3+1][0]*A_r[c3+1][0]
            + A_i[c3+1][0]*A_i[c3+1][0]);
          nf = sqrt(A_r[c3][0] * A_r[c3][0]
            + A_i[c3][0] * A_i[c3][0]);
          sig_r = A_r[c3][0] / nf;
          sig_i = A_i[c3][0] / nf;
          c_r = nf/nm;
          s_r = (sig_r*A_r[c3+1][0] + sig_i*A_i[c3+1][0]) / nm;
          s_i = (sig_i*A_r[c3+1][0] - sig_r*A_i[c3+1][0]) / nm;
          for (c4=0;c4<=N-1;c4++) {
            t1_r = -s_r*A_r[c3][c4] - s_i*A_i[c3][c4]
              + c_r*A_r[c3+1][c4];
            t1_i = -s_r*A_i[c3][c4] + s_i*A_r[c3][c4]
              + c_r*A_i[c3+1][c4];
            t2_r = c_r*A_r[c3][c4] + s_r*A_r[c3+1][c4]
              - s_i*A_i[c3+1][c4];
            t2_i = c_r*A_i[c3][c4] + s_r*A_i[c3+1][c4]
              + s_i*A_r[c3+1][c4];
            A_r[c3][c4] = t1_r;
            A_i[c3][c4] = t1_i;
            A_r[c3+1][c4] = t2_r;
            A_i[c3+1][c4] = t2_i;
          }
        }
      }
    /* And much more */
  \end{lstlisting}

  \caption{Optimized \textsf{Givens} kernel (part)}
  \label{fig:givens_optimized}
\end{figure}

\subsection{Dynamic techniques helping loop transformations}

In the previous experiments, different static methods were always
capable of extracting scalable parallelism. Figure~\ref{fig:gauss}
shows the forward elimination step of the \textsf{Gauss-J} kernel, a
Gauss-Jordan elimination algorithm looking for zero diagonal elements 
at each elimination step. Pivoting is the main source of data-dependent
control flow.

\begin{figure}[h!tb]
  \begin{lstlisting}
for (k=1; k<=n-1; ++k) 
{
  // Make sure that diagonal element is not null
  // 1st data-dependent condition
  if (a[k][k] == 0)
  {
    amax = abs(a[k][k]);
    m = k;
    for (i=k+1; i<=n; i++)
    // Find the row with largest pivot
    for (i=k+1; i<=n; i++) {
      aabs = abs(a[i][k]);
      // 2nd data-dependent condition
      if (aabs > amax) {
         amax = aabs;
         m = i;
      }
    }

    // Row permutation
    // 3rd data-dependent condition
    if (m != k) {
      swap(b[m], b[k]);
      for (j=k; j<=n; j++) {
         swap(a[k][j], a[m][j]);
      }
    }
  }

  // Update a[][]
  for (i=k+1; i<=n; i++) {
    xfac = a[i][k] / a[k][k];
    for (j=k+1; j<=n; j++) {
      a[i][j] = a[i][j] - xfac*a[k][j];
    }
    b[i] = b[i] - xfac*b[k];
  }
}

  \end{lstlisting}

  \caption{Forward reduction step of \textsf{Gauss-J}}
  \label{fig:gauss}
\end{figure}

Just like the \textsf{Givens} rotation kernel, a combination of
skewing and tiling is required to achieve the best performance. Static
analysis alone only extracts parallelism on the intermediate
\texttt{i} loop, leading to a weak $1.5\times$ speedup on 8
cores. Dynamic analysis amounts to speculatively assuming that
diagonal elements are not null, hence that row permutations will be
infrequent. Such a speculative assumption can be used to enable more
aggressive loop transformations: it ``virtually'' eliminates the
dependences due to row permutations, and enables \textsc{Pluto} to
discover the composition of skewing and tiling we were hoping for. The
transformed loop nest follows a similar pattern as \textsf{Givens},
with extra conflict detection code. It may lead to sequential
recomputation of the algorithm on the south-western part of the matrix
defined by an offending diagonal coefficient.

Figure~\ref{fig:speedup_givens_gauss} shows the ideal results
on a $10000\times10000$ random matrix where pivoting is never
required. Coarse grain parallelization and locality optimization
through tiling yield a super-linear $10.54\times$ speedup (both
original and transformed versions are automatically and fully
vectorized).

\begin{figure}[h!tb]
  \begin{center}
    \begin{tabular}{|l||r|r|}
      \cline{2-3}
      \multicolumn{1}{l|}{} &
      \multicolumn{1}{c|}{Static} &
      \multicolumn{1}{c|}{Synergistic} \\
      \cline{2-3}
      \noalign{\vspace{2pt}}    
      \hline
      \textsf{Givens} & 1 & 7.02 \\
      \hline
      \textsf{Gauss-J} & 1.5 & 10.54 \\
      \hline
    \end{tabular}
  \end{center}
  
  \caption{\textsf{Givens} and (ideal) \textsf{Gauss-J} speedups on
    the 8-core target}
  \label{fig:speedup_givens_gauss}
\end{figure}

In practice, the speculation always succeeds in the case of positive
definite matrices.\footnote{A matrix $A$ is positive definite if
  $\mathbf{x}^{T}\!\!A\,\mathbf{x}>0$ for all nonzero $\mathbf{x}$.}
Positive definite matrices have other interesting properties such as
being nonsingular, having its largest element on the diagonal, and
having all positive diagonal elements.  No (partial) pivoting is
necessary for a strictly column diagonally dominant matrix when
performing Gaussian elimination or LU factorization. Fortunately, many
matrices that arise in finite element methods are diagonally dominant:
Figure~\ref{fig:gauss-j-speedups} shows the speedups of Gauss-J
running on different matrices of the Harwell-Boeing collection.
Performance variations are due to the matrix size and misspeculations.

\begin{figure}[h!tb]
  \begin{center}
    \begin{tabular}{|l||r|r|r|}
      \cline{2-4}
      \multicolumn{1}{l|}{} &
      \multicolumn{1}{c|}{Synergistic} &
      \multicolumn{1}{c|}{Misspeculations} &
      \multicolumn{1}{c|}{Size} \\
      \cline{2-4}
      \noalign{\vspace{2pt}}    
      \hline
      \textsf{bcsstruc2}  & 2.72 & 0 & $1806\times1806$ \\
      \hline
      \textsf{watt} & 2.96 & 0 & $1856\times1856$ \\
      \hline
      \textsf{oilgen} & 3.54 & 0 & $2205\times2205$ \\
      \hline
      \textsf{econaus} & 0.83 & 11 & $2529\times2529$ \\
      \hline
      \textsf{psmigr} & 4.46 & 0 & $3140\times3140$ \\
      \hline
      \textsf{gemat11} & 0.88 & 2429 & $4929\times4929$ \\
      \hline
    \end{tabular}
  \end{center}

  \caption{\textsf{Gauss-J} speedups on the 8-core target, with
    different Harwell-Boeing matrices}
  \label{fig:gauss-j-speedups}
\end{figure}

\section{Towards Synergistic Transformations and
  Dynamic Analyses}
\label{sec:synergistic}

The previous experiments show that loop transformations can be very
profitable on dynamic, data-dependent control flow. Sometimes,
conservative results of static analyses are sufficient to enable these
transformations and achieve scalable parallelism. But our point is not
to oppose static and dynamic methods. It is much more interesting to
study the impact of dynamic information on the effectiveness of loop
nest transformations, and to exploit static analysis knowledge to
focus the dynamic analysis effort.

The \textsf{Gauss-J} kernel shows that excellent results can be
expected when operating dynamic analyses (inspection, speculation) and
aggressive loop nest optimizations in synergy.  Indeed, we expect that
the benefits of static and dynamic methods can nurture each other in a
large number of parallelization and loop transformation
problems. Under conservative analysis hypotheses, it may be possible
to transform the control flow to generate more efficient dynamic
analysis code; the result of these analyses may authorize bolder
hypotheses on the dependences (speculative or not), which in turn open
for more aggressive loop transformations.

Our study is still too preliminary to demonstrate the effective
profitability of such a synergistic approach on full
applications. However, it is already possible to sketch the principles
of a polyhedral compilation framework embedding dynamic information
into its search space construction, and generating inspection,
conflict detection and/or recovery code automatically (and on demand).

The polyhedral framework captures three important components of the
semantics of a loop nest in a rich, algebraic framework. These
components are the iteration domains (the set of loop iterations) of
all statements, the access functions for all array references in these
statements, and multidimensional scheduling functions to capture the
relative ordering of the statement iterations. These three components
are represented as systems of affine inequalities (unions of convex
polyhedra). Affine transformations are pushing their way into
production compilers, including GCC \cite{Tri09} and IBM XL,
leveraging two unique advantages:
\begin{itemize}
\item arbitrarily complex compositions of loop transformations can be
  represented, while offering a flexible framework to validate their
  legality \cite{Gir06};
\item well-structured search spaces can be built, allowing the design
  effective heuristics to derive such complex sequences of loop
  transformations automatically, addressing the parallelism and
  locality interplay of modern architectures \cite{Fea92b,Bon08b}.
\end{itemize}

The recent work of Benabderrahmane et al.\ extends the applicability
of the polyhedral framework to data-dependent control flow
\cite{Ben10}, but it still relies on conservative results from static
analysis. There is a clear opportunity to refine the set of affine
dependence constraints defining the search space of affine
transformations. The main challenge is to capture the \emph{outcome}
of the data-dependent condition of an inspection or conflict detection
slice. The condition itself cannot be precisely characterized
statically (otherwise there would be no justification for dynamic
analysis); but it has been generated by a previous compilation pass
that can be designed to retain the causal relation between the outcome
of the condition and the presence of a dependence constraint over a
specific set of statement instances.

For example, considering \textsf{Gauss-J} again, a negative outcome of
the first data-dependent condition \texttt{a[k][k] == 0} guarantees
the absence of any dependence involving the row-swapping
statements. This is the very speculative hypothesis that enabled loop
tiling, improving locality and reducing synchronization overhead.

Since dynamic techniques can also benefit from loop transformations to
become more effective and mitigate their intrinsic overhead, it would
be ideal to derive the inspection, conflict detection or recovery code
from the assumptions made in the polyhedral representation itself. For
example, a parallelization heuristic may choose to weight dependences
according to their likelihood to occur at runtime (based on profile
data), and to ignore some of these dependences when looking for
profitable affine transformations. Once a good candidate composition
of loop transformations is found, the polyhedral code generator
produces not only the transformed (parallel) loop nest, but also the
interleaved dynamic analysis code to validate the original
assumption. This is exactly the principle of hybrid analysis by Rus et
al.\ \cite{rus_hybrid_2003}, but extended beyond parallelism detection
and towards the validation of arbitrarily complex loop nest
transformations.

Considering \textsf{Gauss-J} once again, an expert programmer can
easily guess that the first data-dependent condition has a good
predictability potential on some relevant classes of matrices, and
that the second data-dependent condition \texttt{aabs > amax} is very
unlikely to be a relevant candidate for speculation because it amounts
to precisely predicting the row of the maximum pivot. A compiler
looking for speculative execution points may not be able to figure
this out statically, but it can rely on offline profiling, or
multiversioning and online profiling.  Before opting for a more
expensive speculation strategy, the compiler can leverage static
dependence information to discover that a lightweight inspection
scheme is not sufficient to enable loop tiling: the
permutation-hampering dependences would be detected too late, until
after the completion of the \texttt{i} loop of the update part. In
addition, the compiler can also use static dependence information to
figure out the actual impact of the speculative
hypothesis. Speculating on the negative outcome of the first condition
is sufficient to enable loop tiling, but a highly predictable
condition that does not help refining the dependence constraints is
unlikely to be a good speculation candidate in general. Both
predictability and dependence disambiguation are required: this is
exactly the objective of the sensitivity analysis by Rus et al.\
\cite{rus_sensitivity_2007}, which we would like to revisit in the
context of polyhedral compilation.

\section{Conclusion}

This paper does not attempt to be complete, in terms of
state-of-the-art transformations or dynamic analysis techniques.  Our
goal is to study whether the effectiveness of parallelizing compilers
can or cannot be improved when blending static and dynamic techniques
rather than opposing them. Our findings show that there is a strong
potential in following this path:
\begin{itemize}
\item aggressive loop nest optimizations are required for scalability, and
  it is possible to enable them on data-dependent control-flow;
\item it is possible and profitable to leverage dynamic analysis
  information to enhance the effectiveness and applicability of loop
  transformations.
\end{itemize}
We also sketched how to embed dynamic information into affine
transformation spaces, while synthesizing inspection and/or
speculation code automatically.

We are working on fully automating these techniques. We also plan to
extend parallelism detection among acyclic control-flow regions nested
into loop nests, combining affine loop transformations with decoupled
software pipelining \cite{ottoni_automatic_2005}.

\bibliographystyle{abbrv}
\bibliography{pespma}

\end{document}